\documentclass[conference,a4paper]{IEEEtran}
\IEEEoverridecommandlockouts

\usepackage{cite}
\usepackage{amsmath,amssymb,amsfonts}
\usepackage{graphicx}
\usepackage{textcomp}
\usepackage{xcolor}
\usepackage{algorithm}
\usepackage{algorithmic}
\usepackage{subcaption}
\usepackage{eso-pic}
\renewcommand{\baselinestretch}{0.94}

\begin{document}

\title{RIS-Assisted Secure Transmission with Artificial Noise: Element Allocation and Measurements}

\author{
\IEEEauthorblockN{
Mustafa Furkan Beker\textsuperscript{$\ast$},
Ahmet Muaz Aktas\textsuperscript{$\ast$},
Sefa Kayraklik\textsuperscript{$\ast$},
and Sultangali Arzykulov\textsuperscript{$\circ$}
}

\IEEEauthorblockA{
\textsuperscript{$\ast$}Communications and Signal Processing Research (HISAR) Lab.,
TUBITAK BILGEM, Kocaeli, Turkiye
}

\IEEEauthorblockA{
\textsuperscript{$\circ$}Department of Electrical and Computer Engineering,
School of Engineering and Digital Sciences,\\
Nazarbayev University, Astana, Kazakhstan
}

\IEEEauthorblockA{
mustafafurkanbeker@gmail.com,
\{muaz.aktas, sefa.kayraklik\}@tubitak.gov.tr, sultangali.arzykulov@nu.edu.kz
}
}

\IEEEpubid{\makebox[\columnwidth]{\textbf{979-8-3195-1046-4/26/\$31.00~\textcopyright2026 IEEE}\hfill}
\hspace{\columnsep}\makebox[\columnwidth]{}}

\maketitle

\begin{abstract}
Physical layer security in reconfigurable intelligent surface (RIS)-assisted wireless systems can be improved through coordinated control of signal transmission and RIS configuration. In this work, the base station simultaneously transmits the communication signal (CS) and artificial noise (AN) in the presence of a potential eavesdropper. The RIS is partitioned into two groups of reflecting elements, where a portion enhances the desired CS toward the legitimate receiver, while the remaining elements contribute to AN transmission. Two key parameters govern the system design: a transmit power allocation factor between CS and AN, and an RIS element allocation ratio controlling the partitioning of the reflecting elements. An iterative binary phase optimization strategy is employed to enhance the received signal power at Bob while degrading Eve’s reception. Simulation and experimental results demonstrate that proper joint design significantly improves the achievable secrecy capacity.
\end{abstract}

\begin{IEEEkeywords}
Reconfigurable Intelligent Surface (RIS), Artificial Noise (AN), Physical Layer Security, Secrecy Capacity, RIS Element Allocation, Power Allocation, Experimental Validation
\end{IEEEkeywords}

\section{Introduction}

The broadcast nature of wireless communication systems makes transmitted signals vulnerable to interception by unintended receivers. Conventional security approaches mainly rely on upper-layer cryptographic techniques, which may introduce computational complexity and latency. For this reason, physical layer security (PLS) has attracted significant attention as a complementary method to improve communication confidentiality \cite{1}.  Reconfigurable intelligent surfaces (RISs) have emerged as a promising PLS tool since their passive reflecting elements can be configured to strengthen desired propagation, suppress interference, and improve link reliability \cite{3}. Accordingly, RIS-assisted communication has been widely investigated for physical-layer security \cite{4,5,6}. In parallel, artificial noise (AN) transmission has been widely recognized as an effective PLS technique, where intentional interference signals are generated to degrade the reception capability of potential eavesdroppers while preserving reliable communication for legitimate users \cite{2,6}. 

Several studies have explored RIS-assisted secure transmission. Most existing works focus on either transmit power optimization or RIS phase design \cite{4}. While RIS-assisted secure transmission and AN-aided secrecy enhancement have both been investigated in the literature \cite{5,6}, recent studies have also considered RIS-assisted PLS frameworks that combine AN generation with RIS partitioning and power allocation \cite{8,13}. In these systems, RIS partitioning directly affects the balance between communication signal (CS) enhancement at Bob and AN-assisted degradation at Eve. A closely related AN-driven RIS-assisted PLS system was experimentally investigated in \cite{13}, where fixed CS- and AN-oriented RIS partitions are used and practical RIS phase optimization methods are compared. In contrast, this work treats the RIS element allocation itself as a design variable by defining $\beta=K_b/N$, where $K_b$ denotes the number of Bob-oriented RIS elements. This allows the joint effect of $\alpha$ and $\beta$ to be examined under low-complexity binary RIS phase control \cite{7}, following the cooperative-interference principle for secure transmission \cite{9}.

Motivated by this distinction, this work investigates an RIS-assisted secure transmission framework in which a base station (BS) simultaneously transmits the CS and AN through separately controlled signal paths. The RIS is divided into Bob-oriented and Eve-oriented subsets, and the reflection phases are selected through an iterative binary phase adjustment strategy. The main contributions are threefold: 
(i) a variable-partition RIS-assisted AN transmission scheme is investigated, where $\beta=K_b/N$ controls the number of RIS elements assigned to Bob-oriented CS enhancement and Eve-oriented AN operation, unlike fixed-partition RIS configurations;
(ii) a partition-based iterative binary phase selection method is applied under the practical constraint $\theta_n \in \{0,\pi\}$ to configure Bob- and Eve-oriented RIS subsets according to their assigned roles; 
and (iii) the joint impact of $\alpha$ and $\beta$ is evaluated through simulations and software-defined radio (SDR)-based measurements to analyze the secrecy-capacity trade-off under the considered experimental setup.


\section{System Model}

\begin{figure}[t]
\centering
\includegraphics[width=\linewidth]{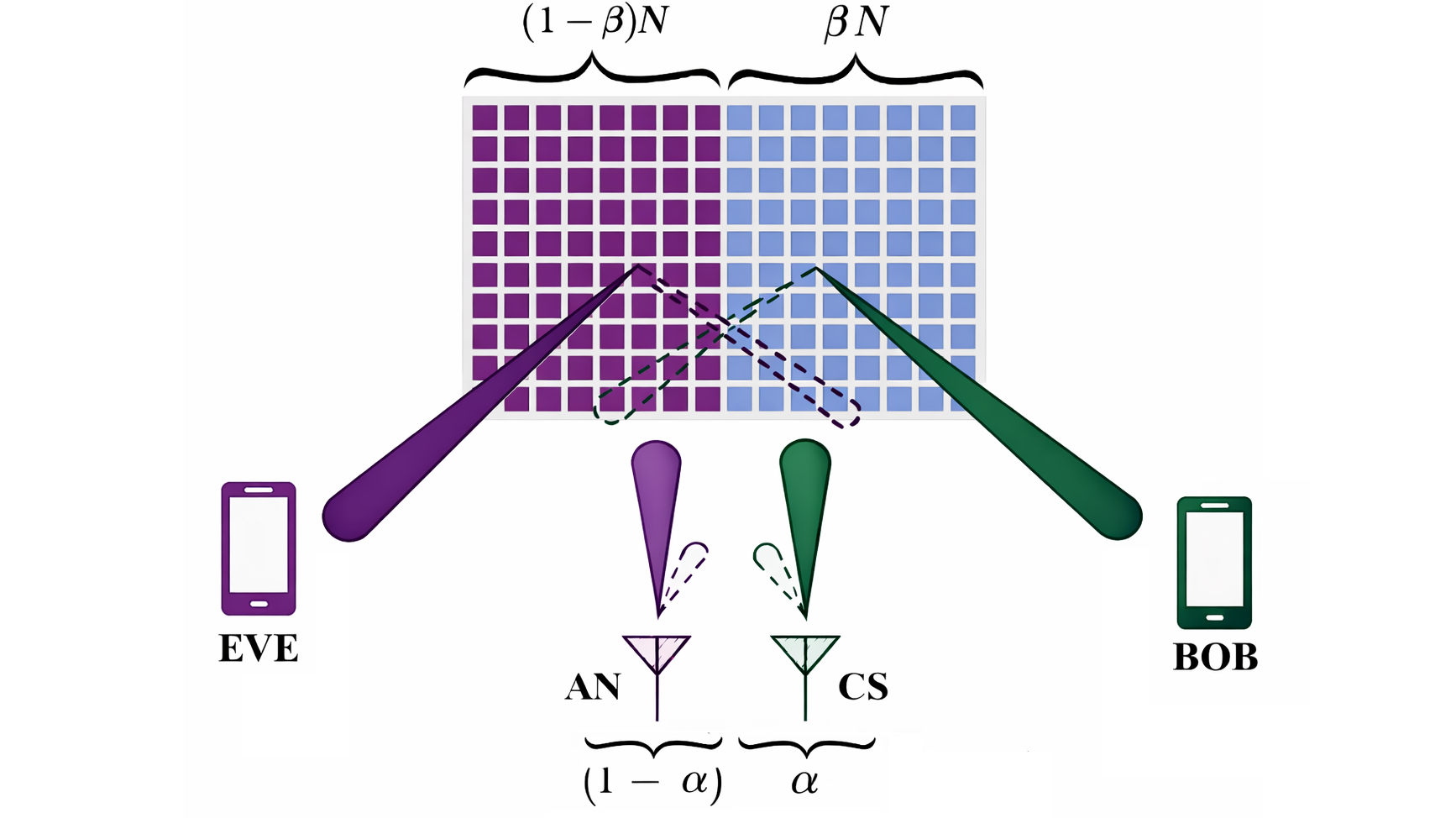}
\caption{System model of the RIS-assisted secure transmission scheme with element allocation and AN generation.}
\label{fig:system_model} \vspace{-12pt}
\end{figure}

An RIS-assisted secure wireless system is considered with a BS, Bob, Eve, and an $N$-element RIS, as shown in Fig.~\ref{fig:system_model}. The BS transmits the CS for Bob and the AN signal to degrade Eve's reception \cite{2,6}.

The RIS is divided into Bob- and Eve-oriented subsets. Let $K_b$ denote the number of Bob-oriented elements and define $\beta=K_b/N$, where $0\leq\beta\leq1$. Thus, $\beta N$ elements mainly enhance the CS link, while $(1-\beta)N$ elements support AN-assisted Eve degradation.

\subsection{Transmit Signal Model}

Let $s$ and $a$ denote the communication and the AN symbols, respectively. Both signals are assumed to have unit average power, i.e., $\mathbb{E}[|s|^2]=1$ and $\mathbb{E}[|a|^2]=1$. The total transmit power is denoted by $P_t$. A power allocation parameter $\alpha$, with $0 \leq \alpha \leq 1$, is used to divide the total transmit power between the CS and the AN. Accordingly, the transmitted CS component can be expressed as $x_s=\sqrt{(\alpha)P_t}\,s$, while the AN component is given by $x_a=\sqrt{(1-\alpha) P_t}\,a$. The CS and AN are transmitted through separate radio-frequency (RF) chains so that they experience distinct BS--RIS channels, enabling independent CS enhancement and AN steering. Therefore, the secrecy performance of the system can be jointly controlled by the transmit power allocation factor $\alpha$ and the RIS element allocation ratio $\beta$.

\subsection{Channel Model}

Let $h_{pn}$ denote the channel coefficient between the transmitter $p$ and the $n$th RIS element, where $p \in \{s,a\}$ represents the CS transmitter and the AN transmitter, respectively. Similarly, let $h_{nu}$ denote the channel coefficient between the $n$th RIS element and the receiver $u$, where $u \in \{b,e\}$ corresponds to Bob and Eve, respectively. These channels are expressed as

\begin{equation}
h_{pn}=|h_{pn}|e^{-j\phi_{pn}},\quad
h_{nu}=|h_{nu}|e^{-j\phi_{nu}}
\end{equation}
where $|h_{pn}|$ and $|h_{nu}|$ denote the channel amplitudes, while $\phi_{pn}$ and $\phi_{nu}$ represent the corresponding channel phases. Each RIS element applies a controllable phase shift $\theta_n$, and the corresponding reflection coefficient is defined as $\varphi_n = e^{-j\theta_n}$. In this work, binary RIS phase control is considered in order to reduce hardware complexity \cite{7}. Accordingly, the RIS phase shifts are restricted to $\theta_n \in \{0,\pi\}$ for $n = 1,2,\dots,N$.

\subsection{RIS-Assisted Cascaded Channel Model}

Since the RIS is partitioned into two subsets, the number of RIS elements assigned to the Bob-oriented partition is given by $K_b=\beta N$, while the remaining $(1-\beta)N$ elements correspond to the Eve-oriented partition. The effective cascaded channel between transmitter $p$ and the receiver $u$ through the RIS can be expressed as

\begin{equation}
G_{p,u}=G_{p,u}^{(b)}+G_{p,u}^{(e)}, 
\quad p\in\{s,a\},\ u\in\{b,e\},
\end{equation}
where the contributions of the two RIS partitions are given by

\begin{equation}
G_{p,u}^{(b)} = \sum_{n=1}^{K_b} h_{pn}\varphi_n h_{nu},
\qquad
G_{p,u}^{(e)} = \sum_{n=K_b+1}^{N} h_{pn}\varphi_n h_{nu}.
\end{equation}

The RIS elements in the Bob-oriented partition primarily enhance the desired signal toward Bob, whereas the RIS elements in the Eve-oriented partition are configured to strengthen the artificial-noise effect toward Eve. Nevertheless, due to wireless propagation, both partitions contribute to the received signals at Bob and Eve.

\subsection{Received Signal Model}

Using the above partitioned cascaded channel definitions, the received signals at Bob and Eve can be expressed as

\begin{equation}
y_u=
\sqrt{\alpha P_t}\,G_{s,u}s+
\sqrt{(1-\alpha) P_t}\,G_{a,u}a+n_u,
\quad u\in\{b,e\}
\label{eq:yu}
\end{equation}
where $n_b \sim \mathcal{CN}(0,\sigma_b^2)$ and $n_e \sim \mathcal{CN}(0,\sigma_e^2)$ denote the additive white Gaussian noise (AWGN) terms at Bob and Eve, respectively.

\subsection{SINR and Achievable Rates}

At Bob, the CS is the desired component, while the AN signal acts as interference. Accordingly, the resulting signal-to-interference-plus-noise ratio (SINR) at the receiver $u$ can be expressed as

\begin{equation}
\gamma_u=
\frac{\alpha P_t |G_{s,u}|^2}
{(1-\alpha) P_t |G_{a,u}|^2+\sigma_u^2},
\quad u\in\{b,e\}.
\label{eq:sinr}
\end{equation}

\begin{algorithm}[t]
\caption{Partition-Based Binary RIS Phase Optimization}
\begin{algorithmic}[1]
\STATE \textbf{Input:} $\alpha$, $\beta$, $N$
\STATE \textbf{Output:} $\boldsymbol{\theta}$ and secrecy capacity $C_s$
\FOR{each $\alpha$}
    \FOR{each $\beta$}
        \STATE Partition RIS into Bob- and Eve-oriented sets
        \STATE $K_b \leftarrow \beta N$, \quad $K_e \leftarrow (1-\beta)N$
        
        \FOR{$n$ in Bob-oriented set}
            \STATE $\theta_n \leftarrow \arg\max_{\theta \in \{0,\pi\}} |G_{s,b}|^2$
        \ENDFOR
        
        \FOR{$n$ in Eve-oriented set}
            \STATE $\theta_n \leftarrow \arg\max_{\theta \in \{0,\pi\}} |G_{a,e}|^2$
        \ENDFOR
        
        \STATE Compute $\mathrm{SINR}_b$, $\mathrm{SINR}_e$
        \STATE Compute $C_s = [C_b - C_e]^+$
        \STATE \textbf{return} $\boldsymbol{\theta}$, $C_s$
    \ENDFOR
\ENDFOR
\end{algorithmic}
\end{algorithm}

These expressions show that $\alpha$ directly controls the power split between the CS and the AN, while $\beta$ affects the cascaded channel gains through the RIS partitioning. The achievable capacities of Bob and Eve are denoted by $C_b$ and $C_e$, respectively, and are defined as

\begin{equation}
C_u=\log_2(1+\gamma_u),\quad u\in\{b,e\}.
\label{eq:rate}
\end{equation}

\begin{figure*}[t]
\centering
\begin{subfigure}{0.46\textwidth}
    \centering
    \includegraphics[width=\linewidth]{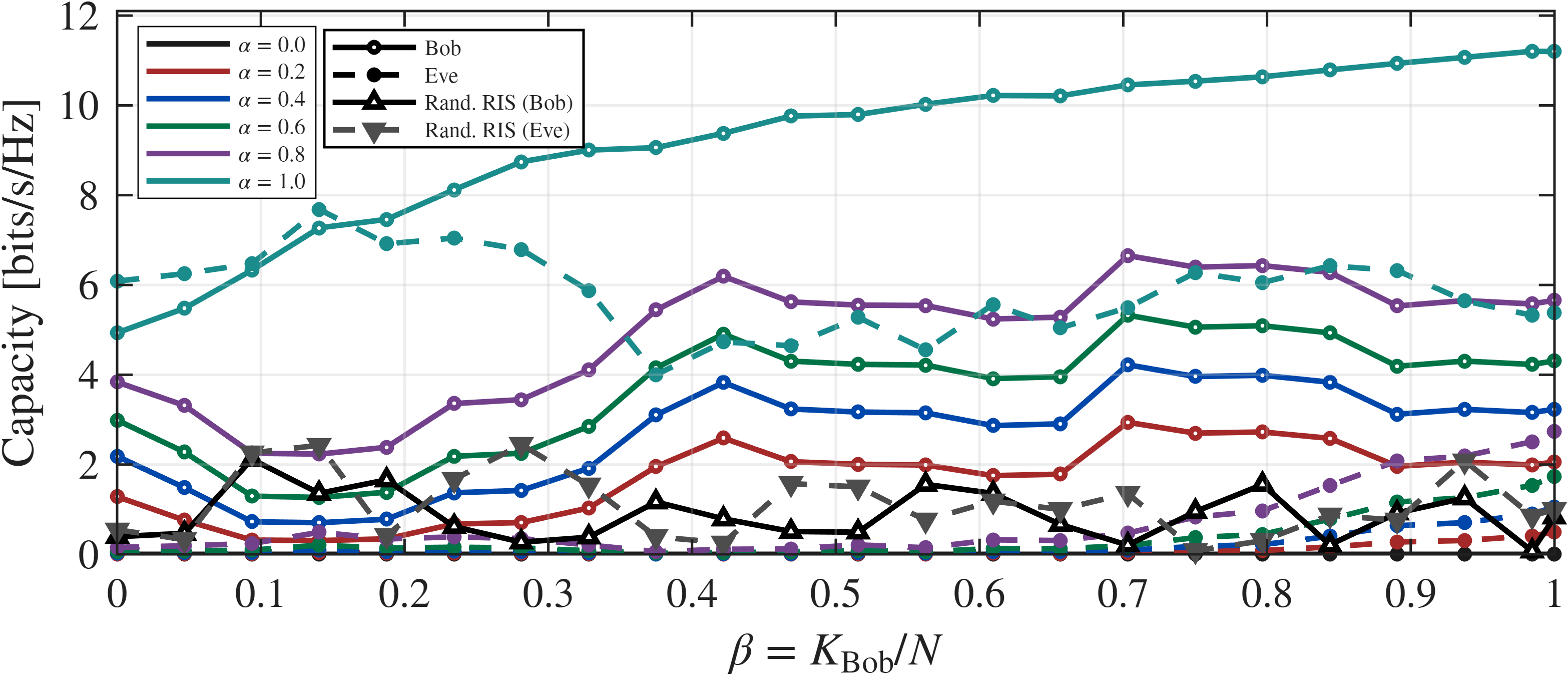}
    \caption{}
\end{subfigure}
\hfill
\begin{subfigure}{0.46\textwidth}
    \centering
    \includegraphics[width=\linewidth]{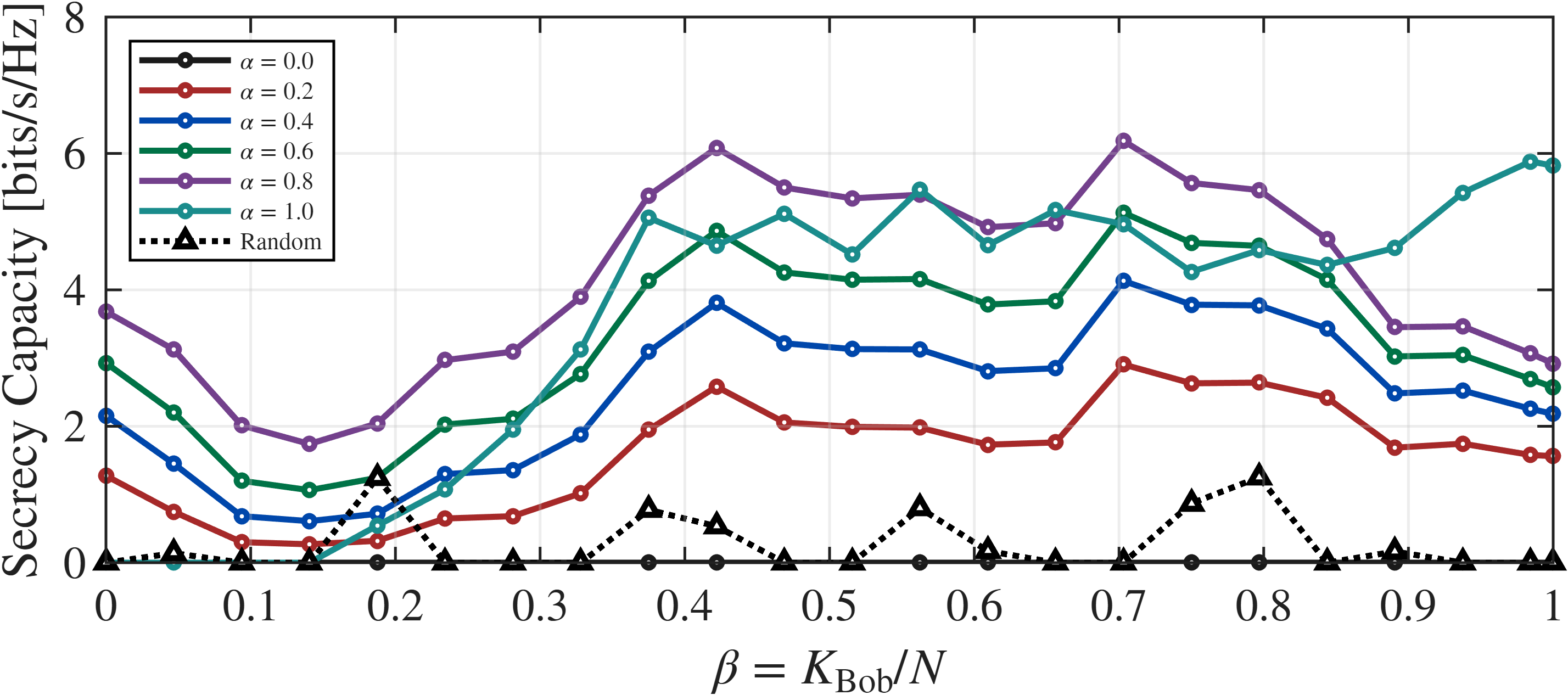}
    \caption{}
\end{subfigure}
\centering
\begin{subfigure}{0.46\textwidth}
    \centering
    \includegraphics[width=\linewidth]{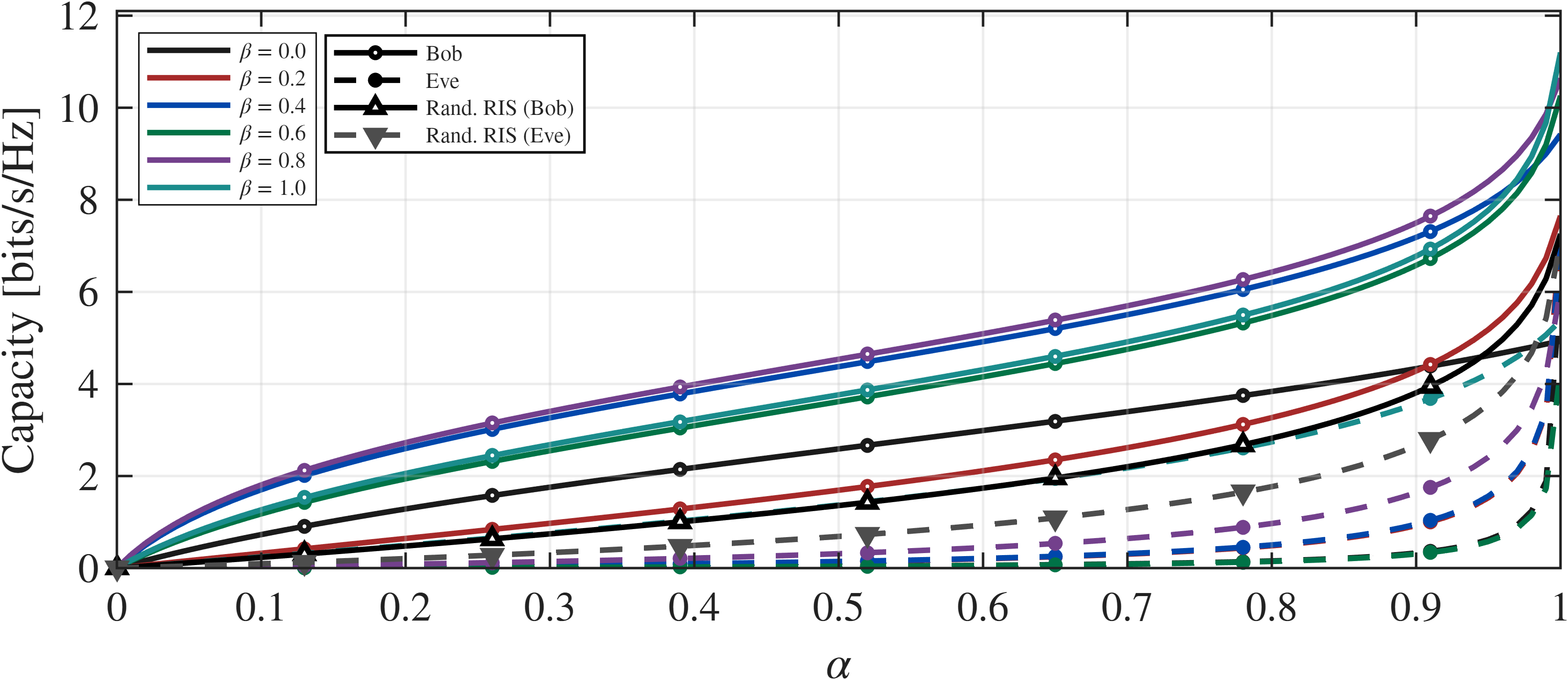}
    \caption{}
\end{subfigure}
\hfill
\begin{subfigure}{0.46\textwidth}
    \centering
    \includegraphics[width=\linewidth]{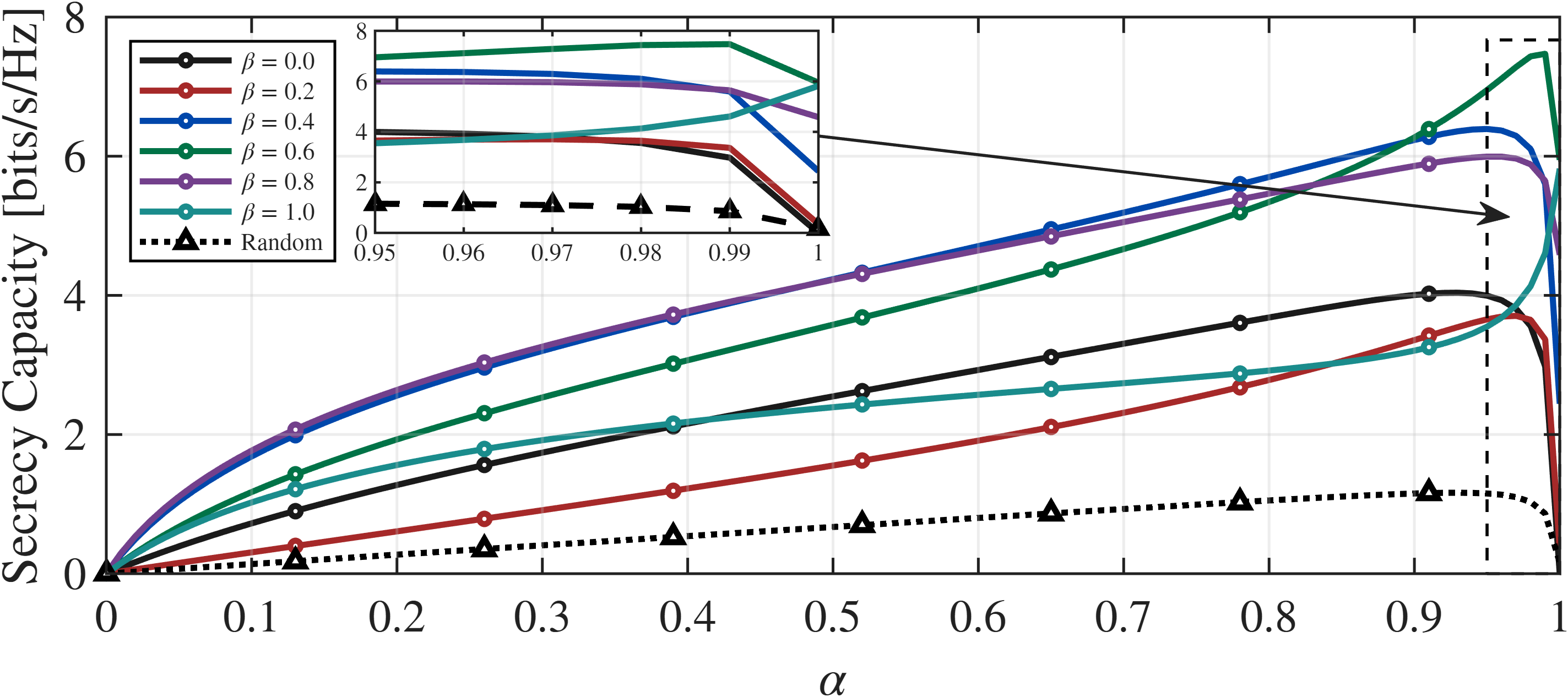}
    \caption{}
\end{subfigure}
\caption{Capacity comparison: (a) $C_b$ and $C_e$ vs. $\beta$; (b) $C_s$ vs. $\beta$; (c) $C_b$ and $C_e$ vs. $\alpha$; (d) $C_s$ vs. $\alpha$.}
\label{fig:performance}\vspace{-10pt}
\end{figure*}

Following the conventional secrecy-theoretic framework of wiretap channels \cite{1}, the secrecy capacity is defined as $C_s=[C_b-C_e]^+$, where $[x]^+ = \max(x,0)$. The secrecy performance can be improved by properly configuring the RIS phase shifts, as well as the transmit power and element allocation. Accordingly, the optimal RIS phase configuration is obtained by maximizing $C_s$ as

\begin{equation}
\begin{aligned}
(\alpha^*, \beta^*, \boldsymbol{\theta}^*) 
&= \arg\max_{\alpha,\beta,\boldsymbol{\theta}} \; C_s(\alpha,\beta,\boldsymbol{\theta}) \\
\text{s.t.} \quad 
&0 \leq \alpha \leq 1,\; 0 \leq \beta \leq 1, &\theta_n \in \{0,\pi\}, \; \forall n.
\end{aligned}
\end{equation}

\subsection{Iterative Binary RIS Phase Optimization}

In the proposed approach, $\alpha$ and $\beta$ are swept over predefined ranges, and for each $(\alpha,\beta)$ pair the RIS is divided into Bob- and Eve-oriented partitions. The Bob-oriented elements are sequentially configured to increase the received CS power at Bob, while the Eve-oriented elements strengthen the AN contribution at Eve. For each RIS element, the two binary phase states $\theta_n\in\{0,\pi\}$ are tested while keeping the other elements fixed, and the phase giving the larger partition-specific objective is selected \cite{iterative}. After the phase updates, the effective cascaded channels, SINR values, $C_b$, $C_e$, and $C_s=[C_b-C_e]^+$ are computed. This process is repeated for all candidate $(\alpha,\beta)$ configurations, and the results are recorded to compare secrecy-performance trends under different power and RIS element allocations.

\section{Simulation and Experimental Results}

The performance of the proposed RIS-assisted secure transmission framework is evaluated in terms of $C_b$, $C_e$, and the resulting $C_s$. The analysis focuses on the impact of the RIS element allocation ratio $\beta$ and the transmit power allocation factor $\alpha$. The simulation parameters are summarized in Table~I.

\begin{table}[t]
\centering
\caption{Simulation parameters}
\label{tab:simulation_parameters}
\renewcommand{\arraystretch}{1.15}
\setlength{\tabcolsep}{6pt}
\begin{tabular}{ll}
\hline
\textbf{Parameter} & \textbf{Description} \\
\hline
Carrier frequency $(f_c)$ & $3.75~\mathrm{GHz}$ \\
Sampling rate $(f_s)$ & $0.5~\mathrm{MHz}$ \\
Transmit power ($P_t$)& $-9\ $dBm\\
Transmit (horn) antenna gain $(G_t)$ & $13~\mathrm{dBi}$ \\
Noise power $(\sigma_{b}^2,\sigma_{e}^2)$ & $-90~\mathrm{dBm}$ \\
Total number of RIS elements $(N)$ & $64~(8 \times 8)$ \\
RIS element spacing $(d_r,d_c)$ & $0.041~\mathrm{m}$ \\
Transmit antenna type & Cosine antenna element \\
RIS cell antenna type & Cosine antenna element \\
Channel model & Free space path loss \\
Alpha sweep range $(\alpha)$ & $0:0.01:1$ \\
Beta sweep range $(K_{\mathrm{Bob}})$ & $0:1:N$ \\
CS transmitter location $(x,y,z)$ & $(0.74,\,0.31,\,0)$ m \\
AN transmitter location $(x,y,z)$ & $(0.74,\,-0.31,\,0)$ m \\
RIS location $(x,y,z)$ & $(0,\,0,\,0.4)$ m \\
Bob location $(x,y,z)$ & $(1.19 , 1.41 , 0)$ m \\
Eve location $(x,y,z)$ & $(1.19 , -1.41 , 0)$ m \\
\hline
\end{tabular}\vspace{-10pt}
\end{table}
Figure~2 presents the simulation results under different RIS allocation and power allocation configurations. In Fig.~2(a), $C_b$ and $C_e$ are shown as functions of $\beta$ for different $\alpha$ values. As $\beta$ increases, more RIS elements are assigned to enhance the CS toward Bob, resulting in a consistent increase in $C_b$. In contrast, $C_e$ remains relatively low for small $\alpha$ values due to the strong AN, but increases as $\alpha$ grows and the AN power decreases.

Figure~2(b) illustrates the corresponding $C_s$ versus $\beta$. Although the boundary case $\alpha=\beta=1$ can yield a high $C_s$ when Eve’s channel is relatively weak, this case corresponds to CS-only transmission and does not actively degrade Eve through AN. Therefore, the effect of $\alpha$ and $\beta$ should be evaluated not only from the maximum $C_s$, but also from the resulting $C_b$--$C_e$ separation. The intermediate allocation region is important because it preserves a high $C_b$ while limiting $C_e$ through AN-assisted RIS operation.

\begin{figure*}[t]
\centering
\begin{minipage}{0.48\textwidth}
\centering
\includegraphics[width=\linewidth]{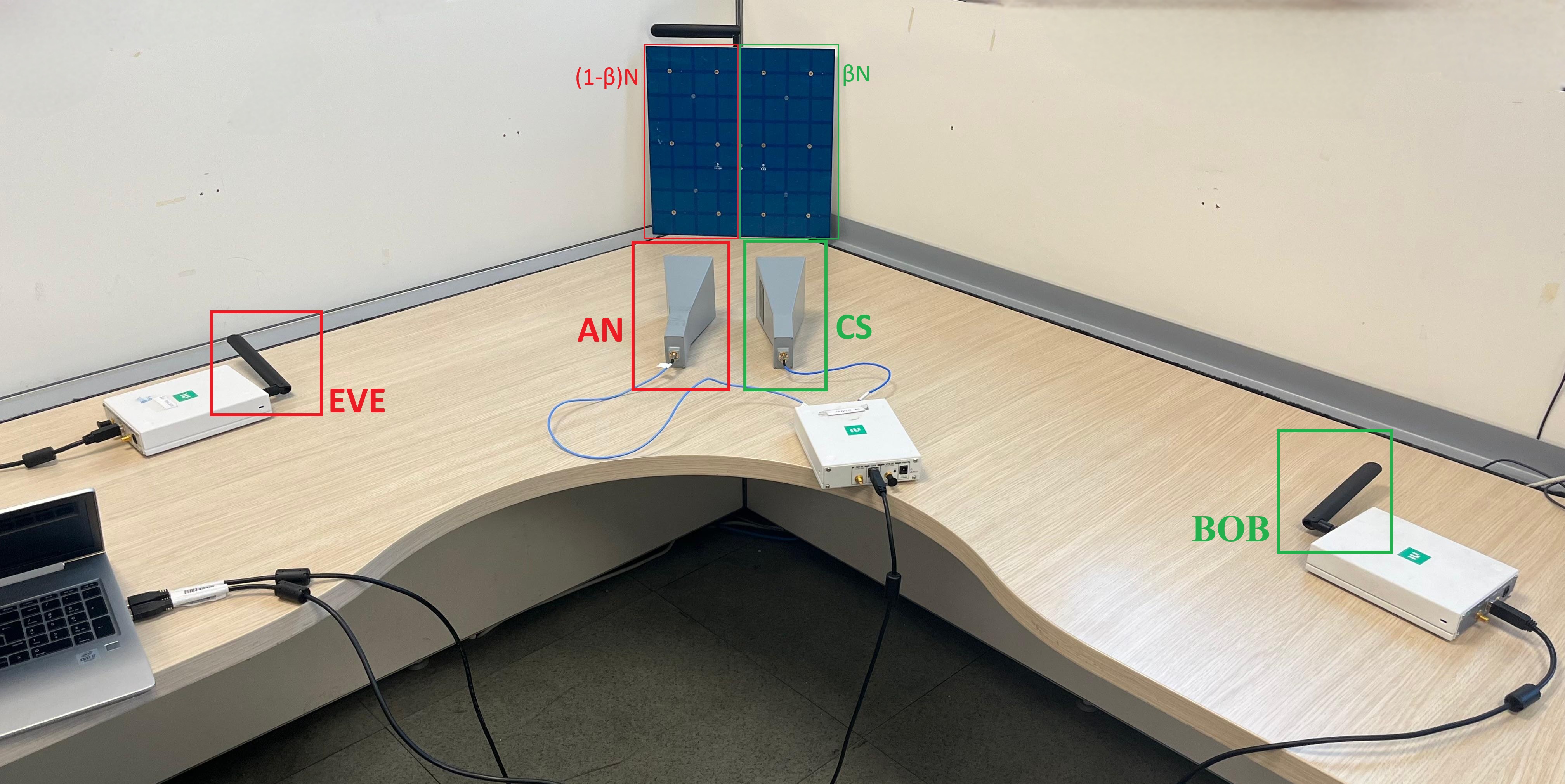}
\caption{Experimental setup of the proposed system.}
\label{fig:exp_setup}
\end{minipage}
\hfill
\begin{minipage}{0.48\textwidth}
\centering
\includegraphics[width=\linewidth]{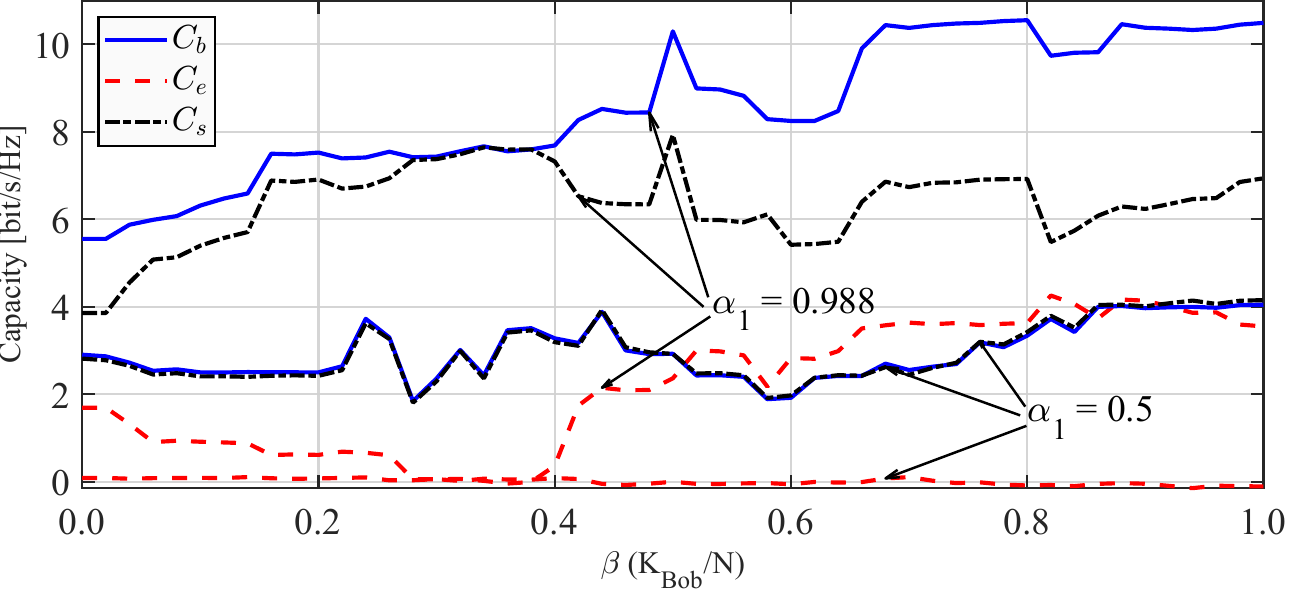}
\caption{Experimental results showing $C_b$, $C_e$, and $C_s$ versus RIS element allocation ratio $\beta$ for different $\alpha$ values.}
\label{fig:exp_results}
\end{minipage}\vspace{-10pt}
\end{figure*}

The impact of $\alpha$ is further examined in Fig.~2(c), where $C_b$ and $C_e$ are plotted as functions of $\alpha$ for fixed $\beta$ values. Increasing $\alpha$ improves $C_b$ due to higher CS power, while simultaneously increasing $C_e$ because of the reduced AN effect. Fig.~2(d) shows the resulting $C_s$ versus $\alpha$, where a non-monotonic behavior is observed. Specifically, $C_s$ initially increases with $\alpha$, reaches a maximum at an intermediate value, and then decreases sharply as $\alpha$ approaches 1. This behavior is due to the near elimination of AN at high $\alpha$, which significantly improves Eve’s reception and degrades secrecy performance. In addition, a random binary phase configuration is considered as a benchmark, where each RIS element is independently assigned a phase from $\{0,\pi\}$ without any optimization. The results show that the proposed iterative phase design significantly outperforms this baseline, indicating that the secrecy gains are mainly due to the optimized RIS configuration rather than random phase assignments.

To validate the practical applicability of the proposed framework, an indoor RIS-assisted SDR-based testbed is considered, following a measurement methodology similar to \cite{yerliRIS}. A single SDR transmitter with two RF chains generates the CS and AN signals, while two receiver nodes emulate Bob and Eve. The RIS is configured using the same element partitioning strategy as in simulations. The experimental setup is illustrated in Fig.~3.

Figure~4 presents the measured $C_b$, $C_e$, and $C_s$ as functions of $\beta$ for different $\alpha$ values. The experimental results are consistent with the simulations. In particular, increasing $\beta$ improves $C_b$ due to RIS-assisted beamforming, while $C_e$ remains limited when sufficient AN power is present. For moderate $\alpha$, a clear gap between $C_b$ and $C_e$ is observed, leading to improved secrecy.

Furthermore, similar to the simulation results, allocating all resources to the CS does not necessarily provide the most meaningful secrecy performance, since it does not actively suppress Eve through AN. Instead, a balanced allocation of RIS elements and transmit power can maintain $C_b$ while limiting $C_e$. Although deviations occur due to indoor reflections, hardware impairments, channel estimation errors, and antenna coupling, the overall trends remain consistent. These results validate the effectiveness of the proposed low-complexity RIS control in practical scenarios.

\section{Conclusion}

This paper investigated an RIS-assisted secure transmission framework that jointly considers transmit power allocation and RIS element partitioning in the presence of AN. Instead of relying on complex optimization techniques, a low-complexity binary phase control approach was employed to configure the RIS, enabling practical implementation.
The results show that the power allocation factor $\alpha$ and the RIS element allocation ratio $\beta$ affect the trade-off between $C_b$ enhancement and $C_e$ suppression. While boundary allocations may provide competitive $C_s$ under favorable channel conditions, balanced allocations are more meaningful for actively degrading Eve through AN-assisted RIS operation. Therefore, the proposed framework highlights the importance of jointly considering $C_s$ and Eve suppression when configuring RIS-assisted secure transmission systems.
Simulation and experimental results exhibited consistent trends, confirming the validity of the proposed approach under practical conditions. Despite hardware impairments and environmental effects, the experimental findings closely followed the simulation behavior, demonstrating the robustness of the framework.
Overall, the study highlights that simple and implementable RIS control strategies, when combined with appropriate resource allocation, can provide significant gains in PLS. These findings offer useful insights for the design of RIS-assisted secure wireless systems in realistic deployment scenarios.

\bibliographystyle{IEEEtran}
\bibliography{references}

\end{document}